\documentclass[12pt,a4paper]{article}
\usepackage{graphicx,amsmath,amsfonts}
\def \ba{\begin{eqnarray}}\def\ea{\end{eqnarray}}
\def\bc{\begin{center}}\def\ec{\end{center}}
\def\nn{\\ \nonumber}
\title{Lepton pair production in relativistic ion collisions to
 all orders in $Z\alpha$ with logarithmic accuracy}
\author{S.~R.~Gevorkyan$^{1,2}$ and E.~A.~Kuraev$^1$}
\begin{document}
\maketitle
\bc
$^{1}$ \it Joint Institute for Nuclear Research, 141980, Dubna, Russia\\
$^{2}$ \it  Yerevan Physics Institute, 375036, Yerevan, Armenia
\ec
\begin{abstract}
 The problem of summation of the perturbation series for the process of
lepton pair production in relativistic ion collisions is investigated.
We show that the  amplitude  of this process can be obtained in the
compact analytical form, if one confines to the terms growing as a powers of
logarithm energy in the cross section, the approximation of
which is completely justified for such  colliders as RHIC and LHC.\\
Using this result we calculate the Coulomb corrections to the cross
section of the process under consideration.
\end{abstract}
The process of lepton pair production in the collision of charged
relativistic particles was investigated many years ago ~\cite{LL,R}
in the lowest order in the fine structure constant $\alpha=\frac{e^2}{4\pi}$.
For the total cross section of the process of lepton pair production in
collision of two relativistic particles with charge numbers $Z_1,Z_2$
\ba
 Z_1+Z_2\to e^+e^-+Z_1+Z_2,
\ea
the famous formula was obtained in 1937 by Racah ~\cite{R}, which in
contemporary symbols reads
\ba \sigma_B=\frac{28}{27}\frac{\alpha^4(Z_1Z_2)^2}{\pi m^2}[L^3-2.2L^2+
3.84L-1.636],\nn
L=\ln(\gamma_1\gamma_2),\gamma_i=\frac{E_i}{m_i},\ea
with m electron mass,$E_i,m_i$ the center of mass energies and masses
of colliding particles.\\
For relativistic heavy ion collisions  this cross section
is huge and the problem of  Coulomb corrections (higher order photon
exchanges between the produced lepton pair and the Coulomb fields of
colliding ions) becomes relevant.\\
The Coulomb corrections (CC) in the case of lepton pair photoproduction off
the nucleus Coulomb field was obtained in ~\cite{BM}.It was shown that
the total cross section for the process of lepton pair
photoproduction in the Coulomb field of the nucleus with charge number Z
reads
\ba
\sigma=\frac{28}{9}\frac{\alpha^3Z^2}{m^2}[\ln\frac{2E_{\gamma}}{m}-
\frac{109}{42}-f(\alpha Z)],f(x)=x^2\sum_{n=1}^{\infty}\frac{1}{n(n^2+x^2)}.
\ea
The production of lepton pairs from the nuclear Coulomb field by virtual
photons was  investigated in~\cite{IM}. In~\cite{KTO}  the Bethe-
Maximon result was obtained in the framework of powerful light-cone
techniques~\cite{BKS}, which allows one to trace carefully approximations
done at its derivation.\\
As to higher order photon exchanges in the process (1), despite the
intensive work done on this issue in the last years ~\cite{BMac,BGMP,
ERG,ISS,LM} the problem of allowing for CC in all orders of perturbation
theory is not solved up to now.The goal of the present work is to
investigate this issue based on our recent results~\cite{BGKN}
obtained for the first terms of  perturbation theory for the amplitude
of the process (1).\\
As was shown in ~\cite{BGKN},  the number of remarkable cancellations between
the different terms in the cross section of the process (1) takes place,
which allows one to simplify the obtained expressions and may be a hint of
how to solve the problem in the general case. In what follows we generalize
the results of~\cite{BGKN} to all orders of perturbation theory and using
the impact factor representation get compact analytical expressions for
 the cross section of the process (1) true to all orders of
perturbation theory (up to the constant terms).\\
It is convenient to separate the full amplitude of the process (1) into
different terms which lead to a distinct energy dependence in the cross
section,namely
\begin{align}
 M &=M_B+M_S+M^S+M_M;\nn
M^S &=\sum_{n=2}^{\infty} M_n^1;M_S=\sum_{n=2}^{\infty} M_1^n;
M_M=\sum_{n,m=2}^{\infty} M_n^m
\end{align}
Here $M_B$ is the Born amplitude (see Fig.1);
the terms $M_S(M^S)$ correspond to the pair production by single photon
exchange between the ion with charge number $Z_1(Z_2)$ and the produced
pair and any multiple photon exchanges between the pair of leptons and the
ion with $Z_2(Z_1)$ (Fig.2); the term $M_M$ in (4) describes the cases
when the pair interacts with each ion more than once (Fig.3).\\
We do not consider the set of Feynman diagrams (FD) accounting for the
Coulomb interactions between the colliding nuclei.Such interactions were
analyzed in ~\cite{BGKN1}, where it has been shown that
interaction between ions completely canceled when the final ions are
not detected.\\
The square of the amplitude (4) is the combination of different terms
which can be classified in the cross section by their energy dependence,
i.e. by different powers of big logarithm $L=\ln(\gamma_1\gamma_2)$.
The Born approximation leads to the third power of L in the total cross
section (see (2)). As was discussed in~\cite{BGKN}, the next powers of L
come from the square of $M_1^n$ and $M_n^1$ and their interference with the
Born amplitude.\\
As to the interference  $M_1^mM_n^1$, it is compensated (up to the constant)
by the interference between the Born term and relevant one  $M_n^m $,
i.e. the combinations of the type $M_B M_n^m$. The same cancellations take
place also in the $\mid M_M\mid ^2$ and any interference  of the type
$M_M M_S$.Each such combination leads in the cross section to the terms
growing with energy as the first power of L.To find their contribution to
the cross section one must take the imaginary part of the complex logarithm
 $\ln (-\frac{s}{m_1m_2}-i\epsilon)=\ln(\frac{s}{m_1m_2}) -i\pi $
which is energy independent.
Thus, if one confines to the terms in the cross section rising with  energy,
the square of the relevant matrix element  can be cast in the following form:
 \ba
 \mid M \mid ^2=\mid M_B\mid ^2+\mid M_S\mid ^2+
 \mid M^S\mid ^2+2Re(M_B^*M_S+M_B^*M^S).
 \ea
Thus to calculate the cross section of the process (1) (up to the constant
terms) one needs the knowledge of the Born amplitude $M_B$ and terms relevant
to single photon exchange with one of the colliding nuclei and any exchanges
with the other $M_S(M^S)$.\\
Let us consider the main results obtained in \cite{BGKN} for the first
terms of such types and generalize them to all orders in $Z\alpha$.
The amplitude of the process (1) in the Born approximation reads ~\cite{BGKN}
\begin{align}
 M_B&=-is\frac{(8\pi\alpha)^2Z_1Z_2N_1N_2(1-\beta_1)(1-\alpha_2)}
{(\vec q_1^2+\beta_1^2m_1^2)(\vec q_2^2+\alpha_2^2m_2^2)}
 \bar u(q_-)R_B v(q_+)\nn
 R_B&=\frac{|\vec q_1|}{s\beta_1}[mR_1\hat e_1+\hat e_1 \hat Q_1+2z_+
 \vec e_1\vec Q+2\mid\vec q_1\mid z_+z_-R_1]\hat p_2;\\
  R_1&=R_1(k_-,k_+)=S(k_-)-S(k_+);\vec{Q}_1=\vec{Q}_1(k_-,k_+)=
 \vec{k}_-S(k_-)+\vec{k}_+S(k_+);\nn
S(k)&=\frac{1}{\vec{k}^2+\mu^2};\mu^2=m^2+z_+z_-\vec q_1^2;
\vec{k}_\pm=\vec{q}_\pm-z_\pm\vec{q}_1;\vec e_1=\frac{\vec q_1}
{\mid \vec q_1\mid}; z_\pm=\frac{\beta_\pm}{\beta_1}
\end{align}
The quantities
$$N_1=\frac{1}{s}\bar{u}(p_1')\hat{p}_2u(p_1);N_2=\frac{1}{s}\bar{u}(p_2')
\hat p_1u(p_2)$$
for collision of spinless  ions $N_1=N_2=1$ so we will omit them later on.\\
We use the Sudakov parameterization for the transferred  momentum $q_1,q_2$
and the created pair momentum $q_+,q_-$ in terms of two light-like 4-vectors
$p_1^2=p_2^2\sim 0 $ built from the momenta of colliding ions and
 invariant energy $ s=(p_1+p_2)^2$, and euclidean two-vectors $\vec q_i$
transverse to a beam direction
$$ q_1=\alpha_1p_2+\beta_1p_1+\vec{q}_1;q_2=\alpha_2p_2+\beta_2p_1+\vec{q},
q_\pm=\alpha_\pm p_2+\beta_\pm p_1+\vec{q}_\pm.$$
For definiteness we suppose that the created pair moves in the direction
of the ion with charge number $Z_1$; thus $\beta_1=\beta _++\beta_-$.
A similar consideration can be easily carried out for the pairs moving in the
opposite direction.\\
Expression (6) depends on lepton transverse momenta $q_\pm$
through  $k_\pm$ unlike the amplitude for the process of pair photoproduction
off the Coulomb field.This is a consequence of  that in
our approach we project all momenta with respect to the axis of colliding
ions in their centre of mass system.As to the case of photoproduction,
the natural choice  is the direction of incident photon.
Finally,as was shown in \cite{KL,BGMS}  the Racah formula (2) can be reproduced
starting with the above expressions.\\
 The spin structure (6) is the same in all orders of perturbation theory
 for the matrix elements relevant to single photon exchange with one of the
 ions (with any number of exchanges with another ion).This is a direct
 consequence of lepton helicity conservation at high energy~\cite{GPS,BGKN}.\\
Using this property we generalize the results obtained in ~\cite{BGKN}
for the first orders in $\alpha$ to all orders of perturbation theory
and represent the amplitude relevant to single photon exchange with the
ion with charge number $Z_1$ and with n attachments to the other (Fig.2):
\begin{align}
M_n^1&=s\frac{(8\pi)^2\alpha Z_1(-i\alpha Z_2)^n(1-\beta_1)}{n!(\vec{q}_1^2+
m_1^2\beta_1^2)}\bar{u}(q_-)\Re V(q_+);\nn
\Re&=\frac{|\vec{q}_1|}{s\beta_1}[mR_n\hat{e}+\hat{e}
\hat Q_n+2z_+\vec{e}\vec Q_n+2|\vec{q}_1|z_+z_-R_n]\hat{p}_2\\
R_n&=\int\frac{d^2k_1...d^2k_n}{\pi^{n-1}\vec{k}_1^2...\vec k_
n^2}\delta^2(\vec k_1+...+\vec k_n-\vec q_2)R_n^1(k_-,k_+,k_1,...k_n),\nn
\vec Q_n&=\int\frac{d^2k_1...d^2k_n}{\pi^{n-1}\vec{k}_1^2...\vec k_
n^2}\delta^2(\vec k_1+...+\vec k_n-\vec q_2)\vec Q_n^1(k_-,k_+,k_1,...k_n).
\end{align}
The quantities $R_n^1(k_-,k_+,k_1,...k_n)\{\vec Q_n^1(k_-,k_+,k_1,...k_n)\}$
can be represented as the combination of S(k) from (7).The expressions for
the cases n=2,3 are cited in ~\cite{BGKN,GI,GPS}.To get the  next orders one
can use the recurrence relations ~\cite{IM}:
\ba
R_n^1(k_n)=R_{n-1}^1(k_-,k_+-k_n)-R_{n-1}^1(k_--k_n,k_+);\nn
\vec{Q}_n^1(k_n)=\vec{Q}_{n-1}^1(k_-,k_+-k_n)-\vec{Q}_{n-1}^1(k_--k_n,k_+).
\ea
These relations read more effectively in terms of the impact parameter
representation which allows one to sum the series like $M_S(M^S)$.
Introducing the Fourier transform of the functions
$R_n(k_n)\{\vec Q_n(k_n)\}$
\ba
G_n^{R\{Q\}}(r_1,r_2)=\frac{1}{(2\pi)^4}\int d^2k_-d^2k_+e^{i\vec{r}_1
\vec{k}_++i\vec{r}_2\vec{k}_-}R_n(k_-,k_+)\{\vec Q_n(k_-,k_+)\}.
\ea
one immediately obtains for the relevant structures
\ba
G_n^R(r_1,r_2)&=&\frac{1}{2(2\pi)^2}K_0(\mu|\vec{r}_1-\vec{r}_2|)\ln^n
\frac{\vec{r}_2^2}{\vec{r}_1^2};\nn
\vec G_n^Q(r_1,r_2)&=&\frac{\mu (\vec r_1-\vec r_2)}{2(2\pi)^2\mid \vec r_1-
\vec r_2\mid}K_1(\mu|\vec r_1-\vec r_2|)\ln^n \frac{\vec{r}_2^2}{\vec{r}_1^2}.
\ea
Here $K_{0,1}(x)$ are the modified Bessel functions of the zero and first order.
Now the summation  of perturbation series can be carried out with the result
\begin{align}
 G_R&=\sum_{n=2}^{\infty}\frac{(-i\nu)^n}{n!}G_n^R(r_1,r_2)=
\frac{K_0(\mu|\vec r_1-\vec r_2|)}{2(2\pi)^2}\sum_{n=2}^{\infty}
\frac{(-i\nu)^n}{n!}\ln^n\frac{\vec{r}_2^2}{\vec{r}_1^2}\nonumber\\
&=\frac{K_0(\mu|\vec{r}_1-\vec{r}_2|)}{2(2\pi)^2}[a^{-i\nu}-1+i\nu\ln a];\\
\vec G_Q&=\frac{\mu(\vec r_1-\vec r_2)K_1(\mu \mid\vec r_1-\vec r_2\mid)}
{2(2\pi)^2\mid{\vec r_1-\vec r_2}\mid}[a^{-i\nu}-1+i\nu\ln a];\nn
\nu &=\alpha Z_2; a=\frac{\vec{r}_2^2}{\vec{r}_1^2}
\end{align}
Applying the inverse Fourier transformation to these expressions one obtains
\begin{align}
R(k_-,k_+)&=\int d^2r_1d^2r_2e^{-i(\vec r_1\vec k_++\vec r_2\vec k_-)}
\frac{K_0(\mu|\vec{r}_1-\vec{r}_2|)}{2}[a^{-i\nu}-1+i\nu\ln a];\\
\vec Q(k_-,k_+)&=\int d^2r_1d^2r_2e^{-i(\vec r_1\vec k_++\vec r_2\vec k_-)}
\frac{\mu(\vec r_1-\vec r_2)}{2\mid \vec r_1-\vec r_2\mid}
K_1(\mu|\vec r_1-\vec r_2|)[a^{-i\nu}-1+i\nu\ln a].
\end{align}
The squares of amplitudes entering  (5)  are determined
by these two structures.Really,as was shown in~\cite{BGKN},
\ba
\frac{1}{4s^2}\sum{\mid\bar u(q_-)\Re v(q_+)\mid}^2=\nn
\frac{1}{2} z_+z_- \vec q_1^2\left\{\left(m^2 + 4z_+^2z_-^2\vec q_1^2\right)
R^2 +(z_+^2+z_-^2)\vec Q^2
+4z_+z_-(z_+-z_-)R \vec q_1\vec Q\right\}.
\ea
As to the amplitude $M_S$ it can be obtained from $M^S$ by simple
replacements~\cite{BGKN}
$$ \vec q_1\to \vec q_2;\vec e_1\to\vec e_2;p_1\to p_2;\beta_1\to\alpha_2.$$
The differential cross section of the process (1) is
connected with the full amplitude (4) by the relation
\ba
d\sigma=\frac{1}{8s}\sum |M|^2 d\Gamma.
\ea
The summation is carried over the final particles polarization.
The phase space volume for four particles in the final state can
be expressed through the variables introduced above and phase
volume of the final leptons
\ba
d\Gamma_{\pm}=\frac{d^3q_+}{2E_+}\frac{d^3q_-}{2E_-}
{\delta}^4(q_1+q_2-q_+-q_-)
\ea
in the following way
\begin{align}
d\Gamma &=\frac{1}{(2\pi)^8}\frac{d^3p_1'}{2E_1'}\frac{d^3p_2'}{2E_2'}
\frac{d^3q_-}{2E_-}\frac{d^3q_+}{2E_+}{\delta}^4(p_1+p_2-p_1'-p_2'-q_+-q_-)
\nn &=\frac{ds_p}{{4(2\pi)^8s(1-\alpha_2)(1-\beta_1)}}\frac{d\beta_1}
{\beta_1}d^2\vec{q_1}d^2\vec{q_2}d\Gamma_{\pm}; s_p=(q_++q_-)^2.
\end{align}
The expressions (5-8,15-20) allows one to calculate the fully differential
cross section of the process under consideration. \\
Let us consider the cross section  integrated over transverse
momenta of final leptons.For this quantity one can obtain the simple
analytical formula from which the special cases can be easily deduce.\\
First of all let us note that the integration over created particles
transverse momenta in the combinations of type $\mid M_S\mid^2+
2Re M_SM_B^*$ leads to the following structures:
\begin{align}
 J_R&=\int d^2r_1d^2r_2 K_0^2(\mu\mid\vec r_1-\vec r_2\mid)[(a^{-i\nu}-1+
i\nu \ln a)(a^{i\nu}-1-i\nu \ln a)\nn
&+i\nu ln a(a^{-i\nu}-1+i\nu ln a)-i\nu\ln a(a^{i\nu}-1-i\nu ln a)]\nn
&=\int d^2r_1d^2r_2 K_0^2(\mu\mid\vec r_1-\vec r_2\mid )[2-a^{i\nu}-
a^{-i\nu}-\nu^2 \ln^2a];\nonumber\\
J_Q&=\int d^2r_1d^2r_2 K_1^2(\mu\mid\vec r_1-\vec r_2\mid)
[2-a^{i\nu}-a^{-i\nu}-\nu^2{ln}^2a].
\end{align}
The further integration leads to the relations ~\cite{IM}
\ba
J_R=-f(\nu)\frac{8{\pi}^2{\nu}^2}{3{\mu}^4};\nn
J_Q=-f(\nu)\frac{16{\pi}^2{\nu}^2}{3{\mu}^4},
\ea
where the function $f(\nu)$ is given by (3).\\
Using the above expressions and representing the cross section
as the sum of Born term and the term relevant to CC contribution
$d\sigma=d\sigma_B-d\sigma_C$ we obtain the following formula:
\ba
d\sigma_C&=&\frac{4\nu_1^2\nu_2^2}{3\pi m^2}(f(\nu_1)+f(\nu_2))\nn
&\times&\frac{[3+4z(1-z)(\frac{q^2}{2m^2}-1)]}{(1+z(1-z)\frac{q^2}{m^2})^2}
\frac{(1-\beta)d\beta}{\beta}\frac{xdx}{(x+\beta^2)^2}dz.
\ea
For the sake of simplicity we consider the case of ions with equal mass
$m_1=m_2=M$ and consequently the same Lorentz factors $\gamma_1=\gamma_2=
\gamma=\frac{E}{M} $.The variables $x=\frac{\vec q^2}{M^2}$ and $\beta$
belongs to the intervals
\ba
0\leq x< \infty;\nn \frac{m}{M\gamma^2}(1+\frac{M}{4m}x)
\leq \beta\leq 1.
\ea
The expression (24) becomes invalid in the case of pair production by the
very soft photons ($\omega\sim 2m $ in the rest system of one of the ion)
which corresponds to the lower limit of $\beta$ ( we are working
in the center of mass system of colliding ions.) At such low energies one
has to take into account the threshold effects and deviation from the
eikonal approach~\cite{AB} which is out of the present work scope.\\
The integrations in (24) can be carried out (for details see Appendix).
We cite here the expression for CC contribution to the total cross section
of the process (1)
\ba
\sigma_C=\frac{28\nu_1^2\nu_2^2}{9\pi m^2}
(f(\nu_1)+f(\nu_2))[\ln^2(\gamma_1\gamma_2)+\frac{20}{21}\ln(\gamma_1
\gamma_2)+O(const)].
\ea
The leading term in this formula $\sim \ln^2(\gamma_1\gamma_2)$ coincides
with the relevant one found in ~\cite{ISS,LM}. As to the next to leading
term ($\sim \ln(\gamma_1\gamma_2)$) in (26) it differs  from relevant one
in ~\cite{LM} (see their formula (17)).The reason is transparent.
We calculated next to leading terms in the part of the cross section
corresponding to the single photon exchange from one of the nuclei
($M_S,M^S$). As to the formula (17) in ~\cite{LM} by derivation it is relevant
to the part of the cross section provided by two or more  exchanges from
both nuclei ($M_M$ with the preceding notation).As was discussed above and
in ~\cite{BGKN} such  exchanges contribute only to the constant part of the
total cross section and are outside of our consideration.There is
correspondence between (26) and relevant formula (30) in ~\cite{IKSS}
obtained for the total cross section of $e^+e^-$ pair production in the
Coulomb field of the nucleus by high energy muons using the generalized
Weizsacker-Williams approximation.\\
In deducing (26) we integrate over $\beta$ from its lower limit,
which is strictly speaking  incorrect by the reasons mentioned above.
Nevertheless  the error in such approach can not be crucial and expression
(26) can be safely used for the simple estimates of the Coulomb corrections
to the total cross section of the process under consideration.The reason is
 relative smallness of the near threshold contribution of the pair
photoproduction in the Coulomb field~\cite{AB}.\\
 The obtained in the present work expressions can be used for calculation
of completely (four particles in the final state) or partly (see (24))
differential cross sections the issue which is necessary in any experiment.
\section*{Acknowledgments}
We are grateful to V.Serbo and A.Tarasov for useful discussions and interest
to the work.We acknowledges the warm hospitality at ECT ,Trento.(E.K) is
grateful to the Institute of Physics SAS (Bratislava), where part of this
work was done.The work is supported by INTAS grants 30494,00366.
\newpage
\section*{Appendix A}
\renewcommand{\theequation}{A.\arabic{equation}}
\setcounter{equation}{0}
Let us calculate the total cross section of the lepton pair production off
the nuclear Coulomb field by virtual photons i.e. the process
\ba
\gamma^*(k)+Z\to e^+(q_+)+e^-(q_-)+Z
\ea
This allow us to compare the results obtained in our approach with the
well known one and show how the integration over phase space can be carried
out.For our purposes it is sufficient to calculate the total cross section
of this process in the Born approximation.The generalization to all orders
in $\alpha$ using our approach is an easy task.\\
Using the same approximations as before, the total cross section of the
process (A.1) in the Born approximation read
\begin{align}
d\sigma &=\frac{2\alpha^3Z^2}{\pi^2(\vec q^2+q_L^2)^2}[(m^2+4z^2(1-z)^2Q^2)
R_S^2+(z^2+(1-z)^2)\vec R_V^2] dzd^2q_+d^2q;\\
R_S &=\frac{1}{\vec q_+^2+\mu^2}-\frac{1}{\vec q_-^2+\mu^2};\vec R_V=\frac
{\vec q_+}{\vec q_+^2+\mu^2}-\frac{\vec q_-}{\vec q_-^2+\mu^2};\nn
\vec q &=\vec q_++\vec q_-;\mu^2=m^2+z(1-z)Q^2;q_L=\frac{\mu^2+\vec q_+^2}
{2z(1-z)\omega}.
\end{align}
Here $Q^2,\omega$ are the mass and energy of virtual photons.
The integration over transverse momenta $ q,q_+$ can be done
analyticaly.We divide the integration over the two-dimensional transfer
momentum q in the two regions: 1)Small momentum transfer where the
longitudinal momentum $q_L$ has to be taken into account $0 \leq q\leq
\xi\sim m $ and 2)the region
where one can safely ignore it,  i.e., $\xi\leq q\leq\infty $.\\
Using this trick the integration over transverse momenta $\vec q,\vec q_+$
can be carried out.As a result,the total cross section for the lepton pair
production by transverse and longitudinal photons in the Coulomb field can
be represented as
\ba
\sigma_T&=&\frac{4\alpha^3Z^2}{3}\int [\ln{\frac{2\omega z(1-z)}{\mu}}-
\frac{1}{2}][(\frac{m}{\mu})^2+2(z^2+(1-z)^2)]\frac{dz}{\mu^2};\\
\sigma_S&=&\frac{16\alpha^3Z^2}{3}Q^2\int{z^2(1-z)^2[\ln{\frac{2\omega z(1-z)}
{\mu^2}}-\frac{1}{2}]\frac{dz}{\mu^4}}.
\ea
The total cross section for the pair production by transversally polarized
photons $\sigma_T$ turns out to  the Bethe-Maximon result (3) at $Q^2$=0.
Moreover, the expression (A.4) is in  agreement with (45) from
~\cite{IM}.As to the pair production by longitudinal photons expressions
(46,59) in~\cite{IM}  have a misprint:the combination $z(1-z)$  enters them
 in the first power in contradiction with our result (A.5).

\newpage
\renewcommand{\textfraction}{0}
\begin{figure}[ht]
\begin{center}
\includegraphics[scale=1.]{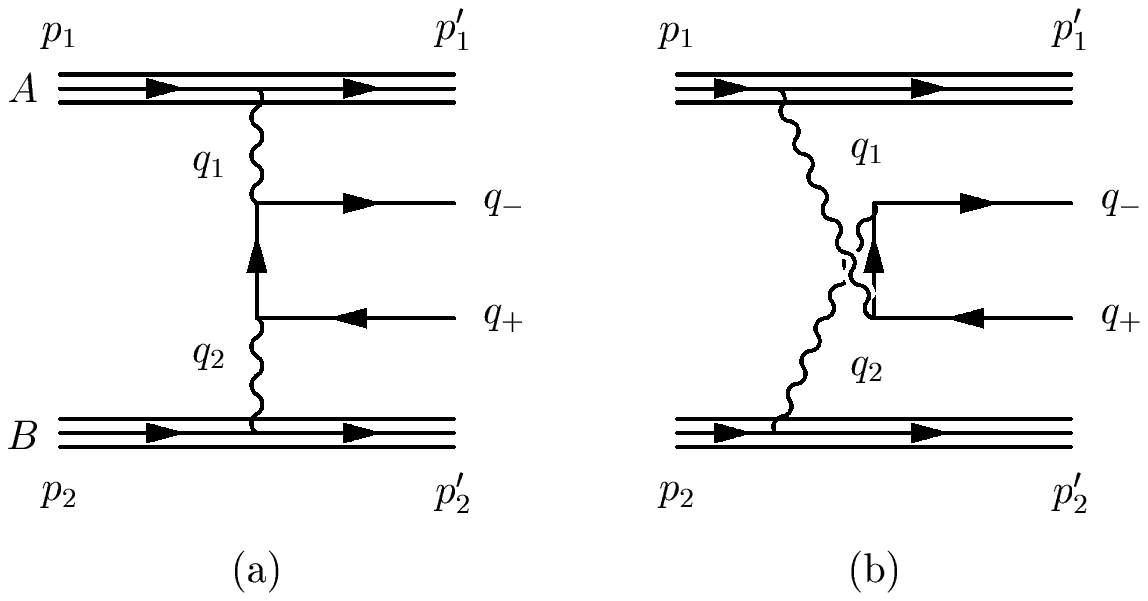}
\caption{The lowest order Feynman diagrams (Born approximation) }
\label{fig:1}
\end{center}
\end{figure}
\begin{figure}[hb]
\begin{center}
\includegraphics[scale=1.]{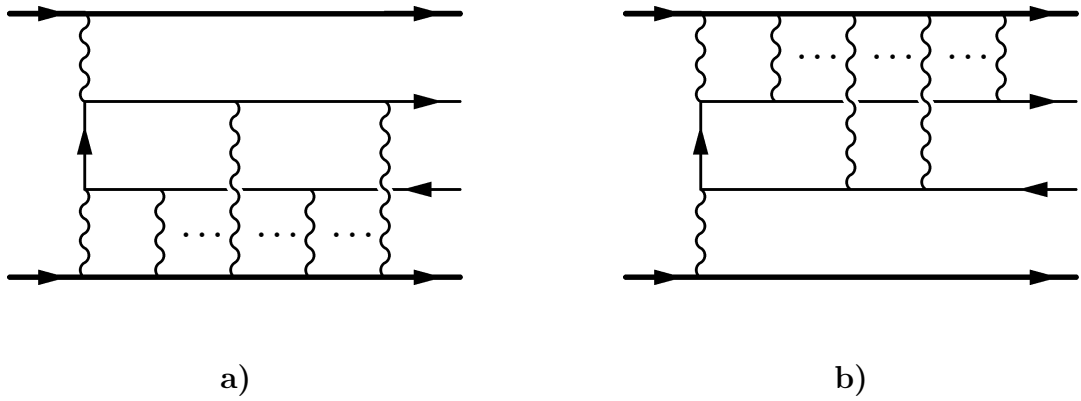}
\caption{The diagrams relevant to the terms with one single
exchange $M_S(M^S)$.}
\label{fig:2}
\end{center}
\end{figure}
\begin{figure}[hb]
\begin{center}
\includegraphics[scale=1.]{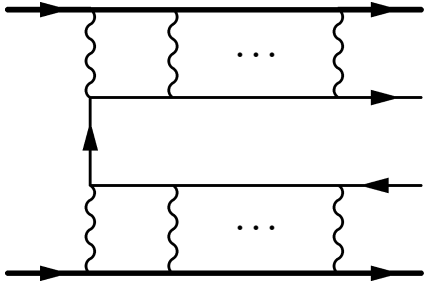}
\caption{Some diagrams describing the amplitudes with multiple exchanges
$M_M$.}
\label{fig:3}
\end{center}
\end{figure}
\end{document}